
%
%
\magnification=1200
\pretolerance=10000
\hsize=4.5 in
\hoffset=1.0 cm
\overfullrule = 0pt
\line{ }
\vskip 0.5 truecm
\rightline {DFTUZ 91.34}
\rightline {July 1992}
\vskip 2. truecm
\centerline{\bf A NEW APPROACH TO
NONCOMPACT LATTICE QED}
\centerline {\bf WITH LIGHT FERMIONS.}
\vskip 2 truecm
\centerline { V.~Azcoiti }
\vskip 0.15 truecm
\centerline {\it Departamento de F\'\i sica Te\'orica, Facultad de
Ciencias, Universidad de Zaragoza,}
\centerline {\it 50009 Zaragoza (Spain)}
\vskip 0.5 truecm
\centerline { G. Di Carlo and A.F. Grillo }
\vskip 0.15 truecm
\centerline {\it Istituto Nazionale di Fisica Nucleare, Laboratori
Nazionali di Frascati,}
\centerline {\it P.O.B. 13 - Frascati (Italy). }
\vskip 1 truecm
\centerline {ABSTRACT}
\vskip 0.3 truecm
\par
We discuss detailed simulations of the non compact abelian model coupled
to light fermions, using a method previously developed that includes
the effects of the fermionic
interactions in an effective action. The approximations involved are related
to an expansion in the flavour number. We address the problem of the
(non) triviality of the theory through a study of
the analytical properties of the effective action as a function of
the pure gauge energy. New numerical results for the plaquette energy,
chiral condensate and a qualitative analysis of the phase diagram are also
presented.
\vfill\eject

\par
\noindent

\item{\bf I.}{\bf Introduction \hfill}
\vskip 1 truecm

The study of
Quantum Electrodynamics on the lattice, i.e. the theory of fermions coupled to
${\cal R}$-valued gauge fields, derives its interest from many reasons. The
obvious one is that its continuum limit (if existing), might describe
standard QED, which is the most succesfull theory at (low energy) perturbative
level.

{}From a more speculative point of view, this theory presents a
challenge  to the
wisdom that only asymptotically free theories are non-trivial in four
dimensions, i.e. the
theory can be defined in the infinite cutoff limit without forcing all the
renormalized coupling constants to zero.

The question is
then posed as whether
the renormalized theory obtained in the limit of infinite cutoff of the
regularized  theory is non trivial.

In recent years, many efforts have been devoted to the study of this problem,
in the context of lattice regularization of the model.

Its compact formulation, not possessing a second order phase transition in
the bare coupling constant [1,2], is not suitable to define a renormalized
continuum theory.

{}From this point of view, the non compact model is much more interesting;
the first numerical investigations of the model, in the quenched
approximation [3], have shown the existence
of a continuous chiral transition at
finite value  of the coupling constant. This transition survives after the
inclusion of dynamical fermions [4-7] so suggesting that the quantum
continuum physics could be reached there.
Moreover, it was believed that the non compact regularization of the
abelian model is in some sense more nearby the continuum formulation than the
compact one.

The theory defined at the critical point as the limit from the broken phase is
interesting by itself, being a theory of strongly interacting fermions, with
spontaneously broken chiral symmetry[8,9]. In this phase the chiral condensate
$<\bar\psi\psi>$ is different from zero in the massless limit.

The interest in the first numerical simulations of the non compact QED
derived from the approximate solution of the Schwinger-Dyson equations
in the quenched
ladder approximation, and the consequent prediction of the existence of
a chiral transition, with an essential singularity scaling law (Miransky
scaling [4,10]), with the aim of testing this result outside the
approximations.

Later,  various groups  have carried out extensive numerical simulations of
this model [4-8,11-15] specifically to determine the critical exponents
and characterize
in this way the nature of the continuum limit, particularly concerning
the issue of the triviality.

The actual situation concerning the determination of the critical exponents,
can be summarized in the following way [14,15]:

\item{i.}Miransky scaling has been disproved, also in the context of
 approximate
numerical solution of the S-D equations [16] .

\item{ii.}Sensible measurements of the critical exponents are extremely
difficult, due to the smallness of the scaling window in which the critical
behaviour can be observed. The only relevant results have been obtained in the
quenched approximation, for large lattices and small fermion masses. These
results contraddict the gaussian character of the fixed point, at least in the
quenched approximation.

An alternative procedure to assess the triviality of the fixed point consists
in
the study of the dependence of the renormalized coupling constant on the
cut-off. In a recent paper [17] we developed such an analysis,
based on general
arguments of block spin Renormalization Group approach and the use of a method
proposed by us [2] to include the effect of dynamical fermions in numerical
simulations of gauge theories.

The aim of this paper is to clarify as much as possible the arguments and
results presented in [17], discussing in detail the fundamental characteristics
of our simulation, the dependence of the results on the flavour number and
fermionic mass and at the same time to present new results for the plaquette
energy and the chiral condensate.

Section II is dedicated to a detailed presentation of our method based on the
definition of an effective fermionic action. We establish a connection between
the functional dependence of the effective fermionic action on the pure
gauge energy and the (non) triviality of the fixed point, and show how a
second order phase transition reflects itself on a non-analyticity of the
effective action.

In Section III we report our results for the effective action, introducing an
expansion in the flavour number, and compare our numerical results with the
analytical predictions developed in the previous section. Sections IV and V
contain our results for the plaquette energy and chiral condensate for various
values of the fermionic mass and flavour numbers. In Section VI we develop a
qualitative analysis of the phase diagram of the model, while in Section VII
we discuss the evaluation of critical indices in the frame of the
Equation of State (EOS) approach. Section VIII contains our conclusions.

\vskip 1 truecm

\item{\bf II.}{\bf The effective fermionic action and triviality \hfill}
\vskip 1 truecm

Addressing the problem of triviality in non compact lattice QED through the
determination of the critical exponents is technically a very difficult
problem: the scaling region is very small and consequently very large
lattices are needed to approach the critical point and characterize
the critical behaviour.
A complete discusson of this is contained in [14,15].

On the other hand in [14] it has been shown that, in the quenched
approximation, critical exponents are definitely distinct from the
ones computed in the Mean Field Theory; it is however not clear wether
this result changes with the inclusion of Dynamical Fermions.

Alternatively, triviality can be studied by computing the renormalized
coupling constant as a function of the cutoff (or equivalently, as a function
of the bare couplings). If the renormalized
coupling constant becomes zero when removing the cutoff the corresponding
fixed point is gaussian.

Two important examples of gaussian fixed point are $\lambda \phi^4$
theory in 4 dimensions and QCD. The fundamental difference between these
two cases is that in the first case it is believed that the renormalized
coupling constant becomes zero at a critical value of the cutoff,
whereas in QCD $\alpha_R$ is zero only in the infinite cutoff limit.
Using the terminology of the Renormalization Group approach, one can say
that in $\lambda \phi^4$, $\lambda$ is an irrelevant coupling,
whereas in QCD $\alpha$ is relevant.

{}From a physical point of view, the main difference between these two models
is that in the first
case $(\lambda \phi^4)$ it is not possible to define a quantum continuum
limit at the gaussian fixed point which is interacting, whereas in the
QCD case this is indeed possible.

Gaussian fixed points are easier to study than non-gaussian ones, since
in the first case one can, in a neighbouring
of the fixed point, perturbatively compute the Callan-Symanzik $\beta$
function, which in turn allows to follow the evolution of the
renormalized coupling constant as a function of the cutoff.

On the other hand, in  Quantum Electrodynamics (QED) and
in perturbation theory, the
renormalized coupling constant is zero, independently on the
value of the bare coupling. This means that the gaussian fixed
point of this model is trivial. Hence the only
possibility of defining a quantum non trivial continuum limit in
this model is that it possesses a non gaussian fixed point, where
weak  coupling arguments do not apply.

The use of non perturbative techniques is therefore essential for
understanding the nature of the continuum limit of the abelian gauge theory.
It is, however, extremely difficult from the numerical point of view to analyse
the fixed point from the study of the critical exponents [14]; instead,
we have performed an analysis of this model based on a indirect study
of the evolution of the renormalized coupling constant.

In this section we will introduce the essential theoretical ideas [17]
at the basis of the numerical results, which will be presented and
discussed in the next sections.

Consider the action of non compact lattice Abelian model coupled to staggered
fermions
\vfill\eject

$$S = {1 \over 2}\sum_{x,\mu}\eta_\mu(x)
{\bar \chi}(x)\{U_\mu(x)\chi(x+\mu)-U_\mu^*(x-\mu)\chi(x-\mu)\}+$$
$$ m \sum_x{\bar \chi}(x)\chi (x) +{\beta\over 2} \sum_{x,\mu < \nu}
F_{\mu \nu}^2(x)\eqno(2.1)$$
$$F_{\mu \nu}(x)=
A_\mu(x) + A_\nu(x+ \hat \mu) - A_\mu(x+\hat \nu) - A_\nu(x) $$

\noindent
where $\beta=1/e^2$ and the fermions are coupled to the
fields $A_\mu(x)$ through the compact link variable $U_\mu(x)=e^{iA_{\mu}(x)}$;
the corresponding partition function is

$${\cal Z} = \int [d \chi] [d \bar \chi] [d A_{\mu}(x)] e^{-S}=
\int [d A_{\mu}(x)] \det \Delta(m,A_{\mu}(x)) e^{-S_G}$$

The main steps of our analysis are:

\item {i.} Determination of an effective fermionic action as a
function of the pure gauge energy, by integrating out all the
other operators of which the effective action is function.

\item {ii.} The existence of a phase transition is monitored through
the appearence of a non analytic behaviour of the effective fermionic
action.

\item {iii.} We then establish a relation between the effective
action as defined in i) and an effective renormalized action. The
(non) linearity of the effective action is then related to the
(non) vanishing of the renormalized coupling constant.

We first define the density of states at fixed pure gauge (non
compact) energy as

$$ N(E) =\int [d{A_\mu(x)}] \delta({1\over 2} \sum_{x,\mu < \nu}
F_{\mu \nu}^2(x) - 6VE) \eqno(2.2)$$

Differently from the compact formulation, the above expression
is divergent even on a finite lattice, owing to the divergence of
the gauge group integration. This problem can be overcome
either by gauge fixing or by factorizing the divergence. In
effect this factorization can be easily accomplished by first
regularizing expression (2.2) multiplying the integrand by a gaussian factor,
corresponding
to the introduction of a mass term for the photon. We define

$$ N(E,M) =\int [d{A_\mu(x)}] \delta({1\over 2} \sum_{x,\mu < \nu}
F_{\mu \nu}^2(x) - 6VE) e^{-M^2 \sum_{x,\mu}{A_\mu(x)}^2} \eqno(2.3)$$

The pure gauge energy ${1\over 2} \sum_{x,\mu < \nu}
F_{\mu \nu}^2(x)$ is a quadratic form, defined through a  real, symmetric
matrix, and can therefore be diagonalized by a unitary transformation.
The number of zero modes of the quadratic form is $V+1$,
so that we can write, in $d$ dimensions,

$$ N(E,M) =\int \prod_{k=1}^{(d-1)V-1}d{B_k} \delta({1\over 2}
\sum_{k}\lambda_k B_k^2-{d(d-1)\over2}VE)$$
$$\times \prod_k e^ {-M^2B^2_k}
\Big[\int dB e^{-M^2B}\Big]^{V+1}
 \eqno(2.4)$$

\par
\noindent
where $\lambda_k, k=1,...(d-1)V-1$ are the non zero eigenvalues of the
quadratic form. The integral in square brackets in the above formula
is  gaussian and contains the whole divergence as $M\to 0$, while the first
factor is finite in the same limit.

The first factor can easily be computed using hyperspherical coordinates
in a $(d-1)V-1$ dimensional space, leading to

$$ N(E)= C_G E^{{(d-1)\over2}V - {3\over 2}} \eqno(2.5)$$

Since the density of states $N(E)$ is known analytically, the partition
function, as expressed in function of the effective fermionic action [2,17],
is now a one-dimensional integral

$${\cal Z} = \int dE N(E) e^{-6 \beta V E} e^{-S_{eff}^F(E,m)} \eqno(2.6)$$

\par
\noindent
where again the divergence of $\cal Z$
is contained in $N(E)$ as a multiplicative
constant $C_G$ of no physical relevance.

The effective fermionic action $S_{eff}^F(E,m)$ in (2.6) is related to the
logarithm of the average value of the fermionic determinant over gauge
configurations of fixed pure gauge energy

$$  e^{-S_{eff}^F(E,m)} =  {\int [dA_{\mu}(x)] \det
\Delta(m, A_{\mu}(x))
\delta({1\over 2} \sum_{x,\mu < \nu} F_{\mu \nu}^2(x) - 6VE)
\over
\int [dA_{\mu}(x)]
\delta({1\over 2} \sum_{x,\mu < \nu} F_{\mu \nu}^2(x) - 6VE)}
\eqno(2.7)$$

Again, numerator and denominator of (2.7) are divergent due to gauge
group volume; however, being the fermionic determinant gauge invariant,
this divergence cancels in the ratio so that $S_{eff}^F(E,m)$ is finite.

The total effective action for this model is therefore

$$ S_{eff}(E,V,\beta,m) =
-{3\over 2} V \ln E + 6 \beta V E + S_{eff}^F(E,m)  \eqno(2.8) $$

\par
\noindent
where we have included the contribution from the density of the states in
the effective action.

The effective fermionic action is linearly divergent with the lattice
volume in the thermodynamical limit; in this limit we can use the saddle point
technique to compute $\cal Z$.

Let assume that this model exibits a second
order phase transition at $(\beta_c,m_c)$. The knowledge of the total
effective action $S_{eff}$ allows, through (2.6), to compute, in principle
exactly, the partition function. What is the manifestation in $S_{eff}^F$
of the phase transition?

Defining $\bar S_{eff}^F(E,m)$ the effective action per unit volume, the
saddle point technique allows to write for the VEV of the mean plaquette
energy and chiral condensate
$< \bar\psi \psi >$ the following expressions

$$ < E_p > = E_0(m,\beta) $$
$$< \bar\psi \psi > = - {\partial\over \partial m} \bar S^F_{eff}(E,m)
\Big|_{E=E_0(m,\beta)}
\eqno (2.9)$$

\par
\noindent
where  $E_0(m,\beta)$ is the minimum of the total effective action at
given $\beta,m$, i.e. the solution of the following  equation

$${1 \over {4E}}-\beta -{1\over 6}{\partial\over
\partial E} \bar S^F_{eff}(E,m)= 0 \eqno (2.10) $$

{}From (2.9), (2.10) above we can derive

$$ C_{\beta} =  {\partial \over \partial \beta} <E_p>=
 - \{{1\over 4E^2_0(m,\beta)}+{1\over6}{\partial^2\over\partial E^2}
\bar S^F_{eff}(E,m)\Big|_{E_0(m,\beta)} \}^{-1}$$
$$ {\partial \over \partial \beta} < \bar \psi \psi >  =  - C_{\beta}
{\partial^2 \over\partial E \partial m}\bar S^F_{eff}(E,m) \Big|_{E_0(m,\beta)}
\eqno (2.11)$$
$$ {\partial  \over \partial m}< \bar \psi \psi > =
- {\partial^2  \over \partial m^2}\bar S^F_{eff}(E,m)\Big|_{E_0(m,\beta)}
+{\partial \over\partial m }E_0(m,\beta){1\over C_{\beta}}
{\partial \over \partial \beta} <\bar\psi \psi> $$

A second order transition implies a discontinuity of the second derivative
of the  free energy. In particular, a discontinuity of the specific
heat can be produced as well by a zero in the
denominator of $C_{\beta}$  given by (2.11) as by
a discontinuity of the second
derivative of the fermionic effective action $ {\partial^2\over \partial
E^2} \bar S_{eff}^F(E,m)$ at the value $E_0(m_c,\beta_c)$
corresponding to the critical values of the parameters $\beta_c,m_c$.
The first possibility, which indeed happens in the large $n_f$ limit, will be
analysed  in detail elsewhere.
In the next Sections we will show that our numerical results for
$\bar S_{eff}^F(E,m)$ in the limit $m\to 0$ strongly suggest the existence
of a non analyticity in the effective fermionic action.

We conclude this section by discussing point iii) on the connection
between the effective fermionic action and the renormalized coupling
constant. Let assume that the continuum limit of the theory is determined by
a gaussian, trivial fixed point. This means that, for a sufficiently large
value of the cutoff, or, equivalently, sufficiently near to the critical
surface, the renormalized coupling constant becomes zero. Therefore
the renormalized action near the critical point will consist only of the
kinetic term of the gauge fields, apart possibly for an additive constant,
i.e.
the total, renormalized effective action defined as in (2.8), near the
critical point, will be linear in the renormalized energy, apart from the
contribution of the density of states proportional to $ \ln E_R$.

Therefore the triviality of the continuum limit can be studied from the
effective fermionic action once its relation with
the renormalized action is known.

This connection can be established in the following way [17]. We first write
the partition function as an integral over the plaquette variables
$F^2_{\mu\nu}$ in the following way

$${\cal Z} = \int [dE_{\mu \nu}(x)] N(E_{\mu \nu}(x)) e^{-S(E_{\mu \nu}(x))}
\eqno(2.12)$$

\par
\noindent
with

$$  e^{-S(E_{\mu \nu}(x))} = $$

$$ {\int [dA_{\mu}(x)]
[d{\bar \chi}(x)][d\chi (x)]\prod
\delta(F_{\mu \nu}^2(x) - E_{\mu \nu}(x)) e^{-S}}
\over
\int [dA_{\mu}(x)]
[d{\bar \chi}(x)][d\chi (x)]\prod
\delta(F_{\mu \nu}^2(x) - E_{\mu \nu}(x))  \eqno(2.13)$$

\par
\noindent
where $S$ in the numerator of (2.13) is the action (2.1) and the
denominator of (2.13) is exactly the density of states $N(E_{\mu\nu}(x))$.
We next apply linear block spin Renormalization Group transformations in
the theory described by the effective action
$S(E_{\mu \nu}(x)) - \ln N(E_{\mu \nu}(x))$ . Our spin variable is the
plaquette
variable $E_{\mu \nu}(x)$ which takes values from $0$ to $\infty$
and blocking
is performed at each $\mu \nu$ plane.

We generate in this way a
series of effective actions $S_R(E_{\mu\nu},N_S)$, where $N_S$ is the number of
blocking steps,
which are equivalent at large distances since
we are integrating out all the short distance details.

The renormalized effective action $S^{eff}_R(E_R)$, defined as in (2.6)
in function of the renormalized energy $E_R={1\over 6V} \sum_{n,\mu<\nu}
E_{\mu\nu}(x)$ can therefore be obtained from action (2.6) by the
substitution $E\to X(m,\beta)E_R$, since it has been
derived through linear block-spin transformations plus a final
linear global transformation.

In particular, apart from the obvious logarithmic contribution from the density
of states, linearity of the renormalized action
is implied by the linearity of the effective
fermionic action.

{}From a practical point of view, the above arguments tell us that the
vanishing of the renormalized coupling constants, as a function of the
bare couplings, can be inferred from the linearity of the effective
fermionic action in the corresponding energy interval.

\vskip 1 truecm
\item{\bf III.}{\bf Computation of the effective action \hfill}
\vskip 1 truecm

The evaluation of the effective fermionic action is obviously not
straightforward, owing to the non locality of the fermionic
determinant, important especially at small masses.

In the abelian, compact case, a discussion of the reliability of this
kind of computation was developed in [2]. In this paragraph this
discussion will be expanded, for the non compact case, in greater
details, particularly concerning the numerical evaluation of the
effective  fermionic action (2.7). To this end, we will expand the
effective fermionic action in powers of the number of flavours, and
discuss the  relative importance of the successive terms of this
expansion.

The effective action (2.7), as stated in the previous section, is related to
the
average of the fermionic determinant, computed over gauge field
configurations at fixed pure gauge energy.

To simplify  the notation, let write this average as

$$e^{-S_{eff}^F(E,m)}= < \det\Delta(m,A_{\mu}(x)) >_E \eqno(3.1)$$

Equation (3.1), using staggered fermions, describes the effective action of
a gauge field coupled to 4 fermion species. In general, for $n_f$ species,
the effective fermionic action will be

$$ e^{-S^F_{eff}(E,m,n_f)} =
<e^{{n_f\over 4} \ln \det \Delta(m,A_{\mu}(x))}>_E
\eqno(3.2)$$

$S^F_{eff}$ can be expanded in cumulants as

$$-S^F_{eff}(E,m,n_f)= {n_f\over 4} <\ln \det \Delta(m,A_{\mu}(x))>_E $$
$$+ {n_f^2\over 32} \{<(\ln \det \Delta)^2>_E-<\ln \det \Delta>_E^2 \}
\eqno(3.3)$$
$$+ {n_f^3\over 384}\{\big<(\ln \det \Delta-<\ln \det\Delta>_E)^3\big>_E\}
+ ... $$

\par
\noindent
which is nothing but an expansion in the flavour number of the effective
fermionic action.

Since the probability distribution of the logarithm of the fermionic
determinant must be computed by numerical methods, the computation of the
successive terms in (3.3) will be increasingly difficult with the order of
the expansion.

In practice, only a few terms will be computed, so that the numerical
results will be affected both by statistical errors, resulting from the
numerical determination of the probability distribution of the logarithm
of the fermionic determinant, and systematic ones,
in consequence of the truncation of expansion (3.3).

Before presenting a detailed discussion of the results for the effective
fermionic action, the relevance of systematic errors will be discussed.

Fig. 1 is a plot of the probability distribution of the logarithm of the
fermionic determinant in a $8^4$ lattice, $m=0$ and normalized pure gauge
energy $E=1.20$. This point has been chosen
because the statistics here is particularly high ($1300$
configurations).
Every gauge configuration diagonalized is separated from
the previous one by $1000$ iterations of a canonical MC process, followed by an
appropriate rescaling of the gauge fields to bring the energy to the required
value.
This procedure guarantees the decorrelation of the successive gauge
configurations that are diagonalized.

The fermionic matrix associated to these gauge configurations is exactly
diagonalized at zero mass through a modified Lanczos algorithm. The
knowledge of all
the eigenvalues of the fermionic matrix at zero mass allows the
computation of the determinant for every value of the mass of the fermions
[2].

Coming back to Fig.1, the continuous line is a gaussian fit of the
distribution measured numerically. The goodness of the fit is evident
($\chi^2/d.o.f.=0.487$) and largely independent from the
fermion mass. If, from these results, we assume that the probability
distribution of the logarithm of the fermionic determinant at fixed pure
gauge energy is gaussian, then only the first two contributions to the
effective fermionic action (3.3) will be different from zero [2] and no
systematic errors will be introduced by the truncation of the expansion.

Moreover, in the thermodynamical limit $V \to \infty$ it is sufficent to this
that the right half (from the maximum) of the distribution is gaussian.

In Figs. 2a,2b,2c we present our results for the first three contributions to
$S_{eff}^F$ respectively, as a function of energy at $m=0$. The results for the
third contribution are compatible with zero, according to the previous
discussion.

Fig. 3 is a plot of the effective action at $m=0$ and 4 flavours computed as
sum of the first two contributions in (3.3); numerical values are also
reported in Table I. From the figure, two different
behaviours of the effective action are evident:
\item{i.} A small energy regime $(E<1)$, typical of the Coulomb phase, where
the effective action shows a linear behaviour as function of the energy.
\item{ii.} A large energy regime, typical of the broken chiral symmetry phase,
$(E>1)$, where the effective action exhibits a non linear behaviour.

In the previous section we have analysed the implications for the effective
action as function of the pure gauge energy of the existence of a second
order transition. Through a saddle point analysis we concluded that a second
order transition should manifest as a discontinuity of the second derivative
with respect to the energy, at a value of the energy corresponding to the
vacuum expectation  value of the plaquette energy at the
critical value of the parameters
$\beta_c, m_c$.

By fitting the experimental points in Fig.3 with two (different) third degree
polynomials, one for $E \le 1.007$ and one for $E\ge 1.007$, one gets very good
fits (continuous line in this figure), with a gap in the second energy
derivative
of the effective fermionic action per unit volume, computed at $E_c$, of
$0.35(5)$. As a result of the fit, we also obtain that the first derivative is
continuous at $E_c$ (notice that a discontinuity in the first energy
derivative should imply from equation (2.10) a first order transition)
and the derivatives of order larger or equal to the
second are zero at $E<E_c$. The value $E_c=1.007(20)$ has been determined
from an analysis of the behaviour of the average plaquette energy, as
discussed in section VI.

The results of this fit, which do not change by changing the order of the
polynomial used for the fit, imply that in effect the second order
transition observed in this model [3-7],  manifests
itself through a non analyticity of the effective fermionic action.
Moreover these results also show
a change of regime from linear to non linear behaviour when passing from
the Coulomb to the chiral symmetry broken phase respectively.

As discussed in the previous section, such a behaviour implies that the
renormalized coupling constant is zero in the Coulomb phase for large
enough values of the cut-off, while it is non zero in the broken phase,
including the infinite cut-off limit. These results do not change qualitatively
 when the number of flavours varies from 1 to 4.

In Table II we report the values of $E_c$, of the gap of the second
derivative of $S^F_{eff}$ at $E_c$ for $n_f = 1,2,3,4$ and of the critical
coupling $\beta_c$, as derived from the average plaquette (see Sect. VI).

To complete this section, we will discuss how these results can be affected by
finite volume effects. Fig. 4 shows the results for the effective fermionic
action per
unit volume at three representative values of the energy
in lattices $4^4, 6^4,
8^4$ and $10^4$. All the points in this figure have been normalized to the
corresponding value in the $10^4$ lattice in such a way that, in absence of
finite volume effects, all the points will lie on a line of constant $(=1)$
ordinate.

The analysis of this figure shows that volume effects in the effective
action decrease rapidly going from $4^4$ to $10^4$.

\vfill\eject
\item {\bf IV.}{\bf Mean Plaquette Energy \hfill}
\vskip 1truecm

The evaluation of the average plaquette energy, as well as that of other
physical observables, is in principle simple in the abelian non compact
model since, once the effective fermionic action is known as function of the
energy, the average plaquette energy can be expressed as the ratio of
one dimensional integrals in $E$.

This simplification derives from the fact that, differently from the compact
case, the density of states $N(E)$ is known analytically in the non compact
model, so that the average plaquette energy can be written as

$$ <E_p> = {\int dE N(E) E e^{-6\beta V E - S^F_{eff}(E,m,n_f)}
\over {\int dE N(E) e^{-6 \beta V E- S^F_{eff}(E,m,n_f)}}} \eqno (4.1)$$

In our case we have measured the effective fermionic action for 28 values of
the energy with the method described in the previous section (see Table I).
Then we have determined the effective fermionic action for $0.3 \le
E \le 0.7$ using a third order polynomial interpolation, and finally we
evaluated numerically the integrals in (4.1).

An alternative method would be to apply directly to (4.1) the saddle point
technique, wich, as well known, is exact in the $V \to \infty$ limit.
 In fact we
have seen that, in a $8^4$ lattice, the two methods give compatible results.

In order to compare our results with others in the literature
and to check
in this way the reliability of our method, we report in Tables III and
IV our results
for  $<E_p>$ for different fermion masses and flavour number (remember that in
our procedure the computations for different masses and flavours are
straightforward and practically no time consuming). Statistical errors
have been computed using standard Jack-Knife procedure.
The agreement with the results reported by other groups [4,6] is extremely
good,
and implies that systematic effects of the method used, as for
plaquette energy
is concerned, are entirely under control.

\vskip 1 truecm

\item{\bf V.} {\bf The chiral condensate \hfill}
\vskip 1 truecm

The vacuum average value of the chiral condensate $<\bar \psi \psi>$ can be
computed as a logarithmic derivative of the partition function (2.6)

$$ <\bar \psi \psi> = - {\int dE e^{-S_{eff}(E,m,\beta,n_f,V)}{\partial \over
\partial m}\bar S^F_{eff}(E,m,n_f)
\over\int dE e^{-S_{eff}}}
\eqno(5.1)$$

\par
\noindent
where we remind that $\bar S^F_{eff} = S^F_{eff}/V$, namely the chiral
condensate is the average value of the derivative with respect to the mass of
the normalized effective action, with a probability distribution deriving from
the total effective action.

The expansion of the effective fermionic action in powers of the number of
flavours (3.3), leads to a similar  expansion for the contributions to the
chiral condensate

$$ - { \partial\bar
 S^F_{eff} \over {\partial m}} = {n_f \over 4} < Tr \Delta^{-1}
>_E +$$
$$+ {n^2_f\over 16} \big< (\ln \det \Delta -<\ln \det \Delta>_E)(Tr
\Delta^{-1}-
<Tr \Delta^{-1}>_E)\big>_E + \eqno(5.2)$$
$$+ {n_f^3\over 64} \big<(\ln \det \Delta-
<\ln \det \Delta>_E)^2(Tr \Delta^{-1}-<Tr\Delta^{-1}>_E)\big>_E+...$$
Therefore the chiral condensate is given by the average value over the
probability distribution defined in (5.1) of the successive terms in (5.2),
normalized by $V$.

Here, as in the computation of the effective fermionic action, the degree of
difficulty in the numerical evaluation of the successive terms in expansion
(5.2) increases with the order in the expansion.

In practice also in this case we will be forced to truncate the expansion to a
certain order, so also the evaluation of the chiral condensate will be in
principle affected by systematic errors due to this approximation.

However, following the analysis done in Section III for the effective fermionic
action, the only non zero contributions to  the chiral condensate are the
first two in (5.2) if the probability distribution of the logarithm of the
determinant at fixed pure gauge energy is gaussian.

Table V contains our results for the chiral condensate at different masses
and flavour number, computed from the first
two contributions to the derivative
with respect to the fermionic mass of the effective action (5.2).
We have not included the contribution proportional to $n_f^3$ since
our results show it compatible with zero.

\vskip 1 truecm
\item{\bf VI.}{\bf The phase diagram \hfill}
\vskip 1 truecm

The numerical results for the effective fermionic action reported in Section
III suggest, as already discussed, the existence of a phase transition
separating a Coulomb phase where the effective action is a linear function
of the energy, from a broken phase characterized by a non linear
behaviour. This picture has been confirmed by the
polynomial
fits to the effective action, which predict a gap in its second derivative at
the critical value of the energy.

To confirm these results and, at the same time, derive a precise determination
of the critical values $\beta_c, E_c$, we will analyse in this section the
numerical results for the average plaquette energy $<E_p>$ as well as its
dependence on $n_f$ and $m$, which will allow to improve our knowledge of the
phase diagram of this model.

Fig. 5 is a plot of our results on the average plaquette energy in function
of $\beta$ in a $8^4$ lattice at $m=0$ and $4$ flavours. The continuous line
in this figure is the plot of $1/4(\beta+h_1(m))$ with $h_1(m)=0.04032$. This
would be the behaviour of the plaquette energy, if the effective action were a
linear function of the energy, with $h_1(m) = {1\over6}
\times$ the slope of the
normalized effective action $\bar S^F_{eff}$ [17].

It follows that the numerical results for $<E_p>$ are to a high degree
consistent with $1/4(\beta+h_1(m))$ in the weak coupling region while in the
strong coupling regime important deviations can be observed.

These results suggest again the existence of a phase transition at an
intermediate $\beta_c$ value of $\beta$. For a precise determination of
$\beta_c$, we present in Fig.6 the dependence of $h_1(m)$ on $\beta$. If the
function $1/4(\beta+h_1(m))$ describes correctly the functional dependence on
$\beta,m$ of the plaquette energy, $h_1(m)$ should be independent on $\beta$.

The results reported in Fig. 6 clearly show the existence of two distinct
regimes in $\beta$ of the behaviour of $h_1(m)$. In the weak coupling region,
i.e. large $\beta$, the results can be fitted with an horizontal
straight line, showing that in fact $h_1(m)$ does not depend on $\beta$.
On the contrary, the region of strong coupling shows a strong dependence of
$h_1(m)$ on $\beta$.

The critical $\beta$ is then defined as the $\beta$ value corresponding to the
intersection of the fits represented by the continuos lines in the figure.
Once $\beta_c$ is known, the evaluation of $E_c$, the critical plaquette
energy, is immediate; this value has been used in Section III to obtain the
polynomial fits of the effective fermionic action.

Table II summarizes our results for $\beta_c$ and $E_c$ in a $8^4$ lattice at
$m=0$ and $n_f=1,2,3,4$.

We next consider the dependence of these results on the bare mass of the
fermion. Fig. 7 shows the behaviour in $\beta$ of $h_1(m)$ for two different
$m$ values, $m=0.0125$ (Fig.7a) and $m=0.1$ (Fig. 7b). The results for
$m=0.0125$ are qualitatively indistinguishable from the results of the massless
case. On the other hand, for $m=0.1$ it is practically impossible to find an
interval in $\beta$ in which $h_1(m)$ is constant, at least in the region
explored in $\beta$.

Our results suggest that the phase transition present at $m=0, \beta=0.208(4)$
continues in the $(\beta,m)$ plane at least up to $m=0.025$. At larger values
of the mass it is extremely difficult to analyse the phase diagram, and,
consequently, to establish if the Coulomb and broken phases are analytically
connected.

In Table VI we also report the critical values $\beta_c,E_c$ at some
representative values of the fermionic mass
for $2$ and $4$ flavours. Fig. 8 presents a tentative
phase diagram in the plane $\beta,m$.

\vskip 1 truecm
\item{\bf VII.}{\bf Critical indices and the Equation of State. \hfill}
\vskip 1 truecm

Although the numerical results reported in [14,15]
show the impossibility of extracting in a meaningful way the critical exponents
from simulations in lattices as small as ours, we think
interesting to make an analysis of the dependence of the value of $\beta_c$
and of the critical exponents on the extrapolation method used to compute the
chiral condensate at zero mass, also in view of the structure of the phase
diagram depicted above.
Notice that our method allows
us to compute the chiral condensate for an arbitrary number of values  of
$\beta,m$  practically
at no extra computer cost, so this analysis is well worth the
effort.

As proposed in [12], we can derive critical indices and $\beta_c$
from the study of the Equation Of State (EOS) which describes the response
of the order parameter of the theory (i.e. $<\bar\psi\psi>$ for QED) to
an explicit (chiral) symmetry breaking term.

In the present case the EOS is (using standard notation [12])

$${<\bar\psi\psi> \over m^{1/\delta}} = F\big({(\beta-\beta_c)\over
<\bar\psi\psi>^{1/\beta_{mag}}} \big)  \eqno(7.1)$$

\noindent
where $F$ is a universal function.
In principle, one could derive from Eq. 7.1 both the critical exponents and
critical coupling.

We have exploited the universal behaviour of the EOS for our data of the
chiral condensate (four flavors) using the expansion (5.2) up to terms
in $n_f^2$.

Owing to the smallness of the lattices used, we have only used data for
$m \ge 0.0125$ for this analysis, for which we believe our data are
fully reliable.
We obtain that our data in this
mass range can be very well fitted by equation (7.1) with
mean field exponents
($\delta=3, \beta_{mag}=0.5$) and $\beta_c= 0.190$ (see Fig. 9). This result
is not surprising since our data for $<\bar\psi\psi>$ are entirely
consistent with the data of ref. [6]

Notice however that our $\beta_c$ obtained from
the behaviour of the plaquette energy, is inconsistent
with equation (7.1). Barring the presence of
a second phase transition different from the chiral one,
our result suggests,
as already stressed in [8] that the determination of
critical coupling and indices from the chiral condensate data in small
lattices gives inconsistent results, since the minimum achievable mass is
relatively high.

On the other hand, our approach to the determination of the chiral condensate
based on expansion (5.2) allows to estimate finite size effects on the various
terms of the expansion. At masses $0.0025\le m \le0.0125$, the coefficient of
$n_f^2$ cannot be reliably evaluated in the lattices we use (it suffers from
strong finite volume effects, the absolute value being decreasing with the
volume).
However, finite
size effects on $Tr \Delta^{-1}$ are small in this mass interval.

We have then decided to investigate the scaling behaviour of the EOS
in the previously mentioned mass interval, in terms
of the approximation to the chiral condensate consisting in the first
term of Eq. (5.2). This approximation should be meaningful, were the critical
behavior of the order parameter contained in this term.
We find scaling behaviour (see Fig. 10) for $\beta_c=0.207$ (a value which is
consistent with the one derived from the behaviour of
the plaquette energy) and
$\delta=2.5, \beta_{mag}=0.64$. On the other hand,
if we impose mean field exponents
in eq. (7.1), it is impossible to find such a good scaling behaviour as that
of Fig. 10 for values
of $\beta_c$ compatible with the one extracted from the
behaviour of the mean plaquette energy.

\vskip 1 truecm
\item{\bf VIII.}{\bf Conclusions \hfill}
\vskip 1 truecm

The first motivation for the work described in the present paper, has been
to test, in a different model, the method proposed in [2] for including the
dynamical effects of light fermions. Further developement of this
investigation has shown the possibility of clarifying some physical phenomena,
in addition to the verification of the reliability of the numerical method.

To this effect, we have presented theoretical and numerical arguments
supporting
the fact that a second order phase transition manifests itself in a non
analyticity of the fermionic effective action as a function of the pure gauge
normalized energy. We have presented numerical evidence for this non analytic
behaviour, which in turn allowed the determination of the values
for the parameters $\beta_c, m_c$ at  the critical point, from the results
for the mean plaquette. This determination is completely independent
from others based on the study of the chiral condensate and, to our knowledge,
it is the first time that the position of the critical point of this model
is determined from the results for the plaquette energy.
The critical values of $\beta$ at $m=0$ obtained at $n_f=2$ and $n_f=4$ with
this method are in perfect agreement with the ones obtained by the Illinois
Group [18].

Using general arguments of the renormalization group approach we have related
the value of the renormalized coupling constants to the functional dependence
of the effective fermionic action on the pure gauge energy $E$. The non
linear behaviour shown by the effective action for energies equal or
larger than a critical energy indicates, in this approach, that some
renormalized coupling constant is different from zero even in the infinite
cut off limit, when we approach it from the broken chiral symmetry phase.

Certainly, these general arguments do not allow to identify the renormalized
coupling (or couplings) which are different from zero in the infinite cutoff
limit [8,9]. Nevertheless, since we define the  critical energy $E_c$, and
consequently coupling $\beta_c$, at the point where the effective action
becomes non linear in the energy,
our results can be reconciled with triviality only
assuming that the phase transition
we have found is different from the chiral one, where the
continuum limit of strongly coupled QED is defined.
We do not see sign
of this behaviour in the lattices we studied.

On the other hand, approaching the critical point from the Coulomb phase, the
linearity observed in the effective fermionic action for energies
characteristic of this phase indicates a trivial continuum limit associated
to the gaussian fixed point.

It is remarkable the fact that, in this phase, the effective fermionic action
has a linear behaviour for all the explored energies. According to the
Renormalization Group arguments we have presented, this implies that
the renormalized coupling constant is zero in this phase for all the
values of the cutoff corresponding to the energies here studied, which
is in qualitative agreement with an analogous phenomenon reported in [13]
and there interpreted as a manifestation of the fact that weakly
coupled QED
becomes an effective weakly renormalizable theory at a small value of the
cutoff.

As for the reliability of the method, we want to stress that the results we
have presented for the plaquette energy $<E_p>$ and chiral
condensate $<\bar\psi\psi>$, are entirely consistent with the results
presented by other groups. This gives evidence to the
accuracy of our method, at least for $n_f\le 4$.
The computational cost of our results is however a small fraction
of that of standard methods [2], fraction which is difficult to
estimate if one takes into
account that the fermionic computation has to be performed only once for
in principle infinitely many
different values of $\beta,m$.

We believe to have presented in this work further evidences supporting the
possibility of a non trivial continuum limit for the strongly coupled
lattice regularized QED. We cannot however exclude that a pathological
behaviour of the theory could make our results not relevant for this limit.

We close by giving some informations about the computing resources used to
produce the results presented in this paper: these informations are useful
to give an idea of the power of the method used.

All these simulations have been run on two four-transputers arrays
of the Theory Group of LNF, with an
estimated total peak computing power of 14.4 Mflops. We have produced with
the microcanonical algorithm and completely diagonalized using a modified
Lanczos algorithm $7771$ configurations in the $4^4$ lattice,
$9370$ in the $6^4$, $6639$ in the $8^4$ and $464$ in the $10^4$
(all fully decorrelated) for a total
of $173.2$ CPU days, roughly corresponding
to  $255$ hours of
a Cray running at $240$ Mflops.

This work has been partly supported through a CICYT (Spain) - INFN
(Italy) collaboration.

\vfill
\eject

\line{}
\centerline {\bf REFERENCES}
\vskip 1 truecm
\item {1.} {J.B. Kogut and E. Dagotto,
 Phys. Rev. Lett. {\bf 59} 617 (1987);
 E. Dagotto and J.B. Kogut,
 Nucl. Phys. {\bf B295[FS2]} 123 (1988).\hfill}
\vskip .1 truecm
\item {2.}
{V. Azcoiti, G. Di Carlo and A.F. Grillo,
 Phys. Rev. Lett. {\bf 65} 2239 (1990);
 V. Azcoiti, A. Cruz, G. Di Carlo, A.F. Grillo and A. Vladikas,
 Phys. Rev. {\bf D43} 3487 (1991).\hfill}
\vskip .1 truecm
\item {3.} {J. Bartholomew, S.H. Shenker, J. Sloan, J.B. Kogut, M. Stone,
 H.W. Wyld, J. Shigmetsu and D.K. Sinclair,
 Nucl. Phys. {\bf B230} 222 (1984).\hfill}
\vskip .1 truecm
\item {4.} {J.B. Kogut, E. Dagotto and A. Kocic,
 Phys. Rev. Lett. {\bf 60} 772 (1988);
 Nucl. Phys. {\bf B317} 253 (1989);
 Nucl. Phys. {\bf B317} 271 (1989).\hfill}
\vskip .1 truecm
\item {5.} {E. Dagotto, A. Kocic and J.B. Kogut,
 Nucl. Phys. {\bf B331} 500 (1990).\hfill}
\vskip .1 truecm
\item {6.}{S.P. Booth, R.D. Kenway and B.J. Pendleton,
 Phys. Lett. {\bf 228B} 115 (1989).\hfill}
\vskip .1 truecm
\item {7.} {M. G\"ockeler, R. Horsley, E. Laermann, P. Rakow, G. Schierholz,
 R. Sommer and U.J. Wiese,
 Nucl. Phys. {\bf B334} 527 (1990).\hfill}
\vskip .1 truecm
\item {8.} {S.J. Hands, J.B. Kogut, R. Renken, A. Kocic, D.K. Sinclair
 and K.C. Wang,
 Phys. Lett. {\bf 261B} 294 (1991).\hfill}
\vskip .1 truecm
\item {9.} {W.A. Bardeen, C.N. Leung and S.T. Love,
 Nucl. Phys. {\bf B273} 649 (1986);
 Nucl. Phys. {\bf B323} 493 (1989).\hfill}
\vskip .1 truecm
\item {10.} {V. Miransky,
 Nuovo Cim. {\bf A90} 149 (1985); P. Fomin, V. Gusynin, V. Miransky
 and Yu. Sitenko, Riv. Nuovo Cim. {\bf 6} 1 (1983).\hfill}
\vskip .1 truecm
\item {11.} {A.M. Horowitz,
 Nucl. Phys. {\bf B17} (Proc. Suppl.) 694 (1990);
 Phys. Lett. {\bf 244B} 306 (1990).\hfill}
\vskip .1 truecm
\item {12.} {A. Kocic, S. Hands, J.B. Kogut and E. Dagotto,
 Nucl. Phys. {\bf B347} 217 (1990).\hfill}
\vskip .1 truecm
\item {13.} {M. G\"ockeler, R. Horsley, P. Rakow, G. Schierholz and R. Sommer,
 Nucl. Phys. {\bf B371} 713 (1992).\hfill}
\vskip .1 truecm
\item {14.} {E. Dagotto, J.B. Kogut and A. Kocic,
 Phys. Rev. {\bf D43} R1763 (1991).\hfill}
\vskip .1 truecm
\item {15.} {A. Kocic, J.B. Kogut, M.P. Lombardo and K.C. Wang, Spectroscopy,
 Scaling and Critical Indices in Strongly Coupled Quenched QED,
 {\bf ILL-(TH)-92-12; CERN-TH.6542/92} (1992).\hfill}
\vskip .1 truecm
\item {16.} {K.I. Kondo, H. Mino and H. Nakatani, Mod. Phys. Lett. {\bf A7}
 1509 (1992).\hfill}
\vskip .1 truecm
\item {17.} {V. Azcoiti, G. Di Carlo and A.F. Grillo, Renormalization Group
 and Triviality in Non Compact Lattice QED with Light Fermions,
 {\bf DFTUZ 91.33} (1991).\hfill}
\vskip .1 truecm
\item {18.} {S.J. Hands, private communication.\hfill}

\vfill
\eject
\line{}
\vskip 1 truecm
\centerline{\bf TABLE CAPTIONS}
\vskip 1 truecm
\item {I}{Effective action and accumulated statistics
versus pure gauge energy, at
$m=0,$ $n_f=4,$ $8^4$ lattice. }

\vskip .1 truecm

\item {II}{ Critical energy, gap of the second
derivative of $S^F_{eff}$ at $E_c$
and critical coupling for $n_f = 1,2,3,4$,
as obtained from the average plaquette data.}

\vskip .1 truecm
\item {III}{Average plaquette energy, $m=0$, $n_f=1...4$. }
\vskip .1 truecm
\item {IV}{Average plaquette energy, $n_f=2$, $m=0.02$, $0.04$ and
$n_f=4$, $m=0.0125$, $0.025$, $0.05$, $0.1$. }
\vskip .1 truecm
\item {V}{Normalized chiral condensate, $n_f=2$, $m=0.02$, $0.04$ and
$n_f=4$, $m=0.0125$, $0.025$, $0.05$, $0.1$. }
\vskip .1 truecm
\item {VI}{Critical energy and coupling for  $n_f=2$, $4$ and
$m>0$.}

\vfill
\eject

\line{}
\vskip 1 truecm
\centerline{\bf FIGURE CAPTIONS}
\vskip 1 truecm
\item {1)}{ Probability distribution of $\log\Delta(m,A_\mu)$ at
$m=0.0$, $E=1.20$, $n_f=4$ and $8^4$ lattice.
The continuous line is a gaussian fit, with $\chi^2/d.o.f.=0.487$ }

\vskip .1 truecm

\item {2)}{ a) first, b) second and c) third cumulant of
$S_f^{eff}$ versus $E$ at $m=0$, $8^4$ lattice.}

\vskip .1 truecm

\item {3)}{ Effective fermionic action in a $8^4$ lattice, $m=0.0$.
Errors are not larger than symbols. The continuous line shows the
two indipendent polynomial fits (see text).}

\vskip .1 truecm

\item {4)}{ Finite volume effects on the normalized effective
fermionic action at three
representative values of the pure gauge energy $E$.}

\vskip .1 truecm

\item {5)}{ Mean plaquette energy versus $\beta$ at $m=0.0$.
The continuous line is a fit of the Coulomb phase data with $E_{pl}=1/4(\beta+
h_1(m))$
and $h_1(0)=0.04032$. }

\vskip .1 truecm

\item {6)}{ $h_1(m)$ versus $\beta$ at $m=0.0$. The solid line in
the strong coupling phase is a polynomial
fit. }

\vskip .1 truecm

\item {7)}{ The same as fig. 6, but for a) $m=0.0125$ and b) $m=0.1$.}

\vskip .1 truecm

\item {8)}{ Tentative phase diagram. The continuous line
represents second order phase transitions. The dashed line corresponds to
values of $m$ where the existence of a phase transition is not clear.}

\vskip .1 truecm

\item {9)}{$<\bar\psi\psi>/m^{1/\delta}$ versus $(\beta_c-\beta)/
<\bar\psi\psi>^{1/\beta_{mag}}$ with $\beta_c=0.190,\delta=3,\beta_{mag}=0.5$
and
$0.0125\le m\le 0.1$, $0.170\le \beta\le 0.215$. Different
symbols correspond to
different mass values. }

\vskip .1 truecm

\item {10)}{ $Tr \Delta^{-1}/m^{1/\delta}$ versus $(\beta_c-\beta)/
(Tr \Delta^{-1})^{1/\beta_{mag}}$ with
$\beta_c=0.207,\delta=2.5,\beta_{mag}=0.64$ and $0.0025\le m \le 0.01$,
$0.170\le \beta\le 0.215$.}

\vfill
\eject

\end